\documentclass[USenglish,oneside,twocolumn]{article}

\usepackage[utf8]{inputenc}
\usepackage[big]{dgruyter_NEW}
\usepackage{xcolor}
\usepackage{xspace}
\usepackage{multicol}
\usepackage{enumitem}
\usepackage[framemethod=tikz]{mdframed}
\usepackage{listings}
\cclogo{\includegraphics{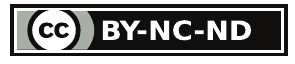}}

\newif{\ifanonymous}
\anonymousfalse

\newcommand{\remove}[1]{}

\newcommand{\AuthorNote}[3]{{\color{#2} {\bf [[ #1:} #3 {\bf ]]}}}

\newcommand{\todo}[1]{{\color{red}\textbf{#1}}}
\newmdenv{licensetext_}
\newenvironment{licensetext}{\begin{licensetext_}\raggedright\it}{\end{licensetext_}}

\newenvironment{myindentpar}[1]%
 {\begin{list}{}%
         {\setlength{\leftmargin}{#1}}%
         \item[]%
 }
 {\end{list}}

\begin{document}
 
\ifanonymous
\else
  \author*[1]{Micah Altman}

  \author[2]{Stephen Chong}

  \author[3]{Alexandra Wood}

  \affil[1]{Program on Information Science, MIT Libraries, E-mail: escience@mit.edu}

  \affil[2]{Harvard John A. Paulson School of Engineering and Applied Sciences, E-mail: chong@seas.harvard.edu}

  \affil[3]{Berkman Klein Center for Internet \& Society, Harvard University, E-mail: awood@cyber.law.harvard.edu}
\fi

  \title{Formalizing Privacy Laws for License Generation and Data Repository Decision Automation}

  \runningtitle{Formalizing Privacy Laws for License Generation and Data Repository Decision Automation}

\abstract{
Legal and regulatory requirements governing transfers of potentially sensitive or private data about individuals are complex and often not very well understood by researchers managing human subjects data. Because there is no common set of licensing agreements that is shared across institutions and laws, current methods to create data use agreements typically require significant efforts by legal counsel for each data transfer. As a result, different institutions duplicate effort in developing variations of agreements for identical uses, and the licenses produced are often too generic and fail to accurately capture either the necessary restrictions on data use or the necessary protections. 
In this paper, we summarize work-in-progress on expert system support to automate some data deposit and release decisions within a data repository, and to generate custom license agreements for those data transfers. 
Our approach formalizes via a logic programming language the privacy-relevant aspects of laws, regulations, and best practices, supported by legal analysis documented in legal memoranda. This formalization enables automated reasoning about the conditions under which a repository can transfer data, through interrogation of users, and the application of formal rules to the facts obtained from users. The proposed system takes the specific conditions for a given data release and produces a custom data use agreement that accurately captures the relevant restrictions on data use. This enables appropriate decisions and accurate licenses, while removing the bottleneck of lawyer effort per data transfer. The operation of the system aims to be transparent, in the sense that administrators, lawyers, institutional review boards, and other interested parties can evaluate the legal reasoning and interpretation embodied in the formalization, and the specific rationale for a decision to accept or release a particular dataset.
}

  \journalname{Proceedings on Privacy Enhancing Technologies}
  \startpage{1}

  \journalyear{2020}
  \journalvolume{2020}
  \journalissue{2}

\maketitle
\section{Introduction}
\subsection{Motivation: Sharing Sensitive Data Governed by Privacy Laws}

Making data widely available supports the replication of research results, increases access to public resources, leverages investments in generating research data, and advances research and innovation \cite{Borgman2013,10.1371/journal.pcbi.1003542}. Moreover, funding agencies often require that data be made openly available where possible (e.g., \cite{nih-sharing,bls-surveys}), and many scholarly journals expect or mandate the sharing of data along with publication \cite{PiwowarC2008}. However, a major challenge for disseminating data is protecting the privacy of human subjects \cite{nrc-expanding-access2005,nrc-people-on-map2005,nrc-beyond-hipaa2009,nrc-biosocial-survey2010}. A complex array of laws, policies, and agreements restrict how data containing personal information may be collected, analyzed, and shared in different contexts.

Privacy laws are highly sector- and context-specific, and contracts for sharing different types of information in different jurisdictions vary significantly. For example, a researcher who conducts a study of human subjects must consider whether the Common Rule (45 C.F.R. part 46) requires approval from an institutional review board and delineates the procedure for obtaining informed consent from subjects. A researcher may also be required to determine whether sharing of the data is governed by the Health Insurance Portability and Accountability Act (HIPAA) Privacy Rule (45 C.F.R. part 160 and subparts A and E of part 164), which regulates the sharing of data containing protected health information held by covered entities, or the Family Educational Rights and Privacy Act (20 U.S.C. § 1232g), which regulates the sharing of records held by educational agencies and institutions that directly relate to a student.

Determining which provisions of the various privacy laws in place govern a given data transfer, and how their requirements should be reflected in a data use agreement, can require a complex legal analysis. The complexity increases when multiple laws, such as a federal privacy law and a state data security law, as well as multiple institutional policies, may be applicable. Such determinations require a thorough analysis of the legal definitions used, the categories of data and entities covered, and the interpretations of these terms by the courts, agencies, and other institutions. In addition, data are often collected or shared under binding data sharing agreements or institutional policies. Researchers and institutions that manage and share data related to individuals are required to understand whether any of these regulatory and contractual restrictions apply to an individual dataset and, if so, how such restrictions affect the handling of the dataset throughout its lifecycle.

Our approach renders the challenges of automating data sharing decisions constrained by complex regulatory requirements are made tractable by limiting the domain of potential action in two key ways: (1) through a selection of use cases and (2) a specification of conditions under which actions are permitted or denied, within the scope of the use cases identified.

First, the scope of analysis is limited to a specific domain of use cases, actors, and actions. In developing our system, we chose to focus on the data sharing decisions made by a research data repository as our initial use case. 

By focusing on a relatively narrow scope of action, this approach aims to overcome many of the challenges associated with formalizing legal requirements. Such challenges include the need to account for ambiguity and flexibility in interpretation, the broad applicability of regulatory requirements, the relevance of multiple laws and policies to a given setting, and variations in practices and policies for complying with legal requirements across institutions. 

Second, the scope of analysis in this formalization is also limited to a subset of conditions that can be automatically permitted or denied. The formalization aims to model only the set of conditions under which actions are clearly permitted or denied. In less clear cases, where automation is not feasible or desirable, the system supports the possibility of escalation to human review.

\subsection{An Illustration: Sharing Educational Data for Research}

Consider a scenario in which a professor at a research university is a principal investigator (PI) leading a ten-year longitudinal education study funded by her university and a private foundation. The research subjects are public school students in grades 6--12. Their parents received notice of the study and provided written consent for their children's participation. The dataset from the study contains grades collected pursuant to transcript requests to the school (with authorization from the students' parents), and responses to questionnaires distributed to the students. Some identifiable information such as names and addresses have been removed, but other information such as gender, race, and ZIP code remain.

At the end of the study, the PI wishes to make the dataset available to other researchers for replication and secondary research uses, as required by the policy of the private foundation that funded the study. She is aware that the sensitive data from this study is likely protected by laws such as FERPA, but she is not very familiar with specific legal requirements, such as the steps that must be taken to de-identify data in accordance with FERPA, or who the data may be permissibly shared with and under what license terms. Furthermore, she does not have the resources to vet individual requests for the data nor apply state-of-the-art de-identification techniques to the data. She wishes to transfer the dataset---and these responsibilities---to a research data repository. She would also like to do so in a way that demonstrates accountability to key stakeholders, including her university's and future researchers' institutional review boards.

She chooses a data repository housed at another research university that serves a worldwide community of researchers from a wide range of disciplines. The repository, in turn, aims to facilitate data access and provide persistence for research data, while managing legal and ethical risks related to the long-term storage of personal data about research subjects. However, the repository also has concerns about accepting personal data that would create new responsibilities for the repository under the law, including the imposition of new administrative, physical, and technical safeguards and audit and retention requirements. The repository is also concerned about accountability to the depositing researcher's home institution for the legal risks it assumes, and about following best practices for exposing research subjects to data privacy risks when storing and transferring data about them.

It is a persistent challenge to ensure that each of the stakeholders involved---including the research subjects, the researchers who collected data about them, the researchers who will use the data, the researchers' home institutions, their institutional review boards, the data repository, funders, and journals---understand the legal restrictions and best practices that apply to the data, and to demonstrate that such restrictions and best practices are being followed to protect the privacy of the individuals in the data.

\subsection{Our Approach}

Our approach aims to make it easier for data producers, users, and publishers to store, use, and transfer sensitive personal information in a standardized and responsible way. We focus on research data repositories, which enables us to understand the challenges and benefits of implementing an automated solution, and the  activities and concerns of research data repositories are shared by other institutions, such as government agencies and commercial firms. (See \citep{altman2018practical, altman2015towards} for an introduction to the laws and risks related to storing and releasing personal data in the corporate, research, and government sectors.) 

We aim to create legal instruments that consistently apply privacy protections across the information lifecycle (from collection to storage to use to release), are modular and machine-actionable, and are tailored to the specific needs of researchers, institutional adopters (e.g., institutional review boards, university offices of sponsored programs, repository owners), and research subjects, for the purposes of ensuring privacy, security, consent, and accountability, and for meeting legal requirements and the users' goals for analysis of the data. To this end, this effort brings together expertise from computer science (formal policy design), law, social science (survey design), and information science (taxonomies).

In our system, we encode the relevant provisions of \emph{domains} (e.g., laws, policies, contracts, and best practices)
as logic programs that can  determine which restrictions are appropriate for a given dataset.
Domains may be defined by legislation, regulation, case-law, or practice. They are defined by humans and human institutions, and thus the boundaries are inherently fuzzy. We rely on legal experts to define the boundaries of each domain by identifying the set of documents (legislative code, commentary, institutional policies, etc.) that must be understood  to act correctly within that domain. 

We then use automated interviews to elicit from users relevant information (such as properties of a dataset). Using these facts and the logic program, we can determine appropriate data handling policies and generate custom license agreements.

 The creation of the logic program, interview questions, and license terms, process is given structure by the application of methodology from law, information science, and computer science. This process, outlined in Table~\ref{formprocess} involves two stages. In the first stage, we conduct use-case analysis of the data curation lifecycle to identify the scenarios in which the repository interacts with other actors over data. In the second stage, we conduct a legal domain analysis, which aims to characterize the restrictions that specified legal domain places over the scenarios identified in the use case analysis.

\begin{table*}[htb]
\centering
\begin{tabular}{p{0.22\linewidth} | p{0.24\linewidth} | p{0.2\linewidth} | p{0.25\linewidth}}
  1. Use Case Analysis: 
  A. Define  Cases in Scope
  \textnormal{
  \begin{enumerate}[label=\alph*.]
  \item Data Deposit
  \item Retention
  \item Transformation
    \item Dissemination
  \end{enumerate}
  }
  &

  2. Use Case Analysis: 
  B. Parse Components
  \textnormal{
  \begin{enumerate}[label=\alph*.]
  \item Actors: Data Controller, Repository, Data Depositor, Data User, Data Subject
  \item Actions: Deposit, Accept, Store, Release, Analyze
  \item Objects: Data set, Record, Consent Form, DUA
  \end{enumerate}
}

  &
  2. Domain Analysis
  B. Legal Characterization
  \textnormal{
  \begin{enumerate}[label=\alph*.]
  \item Identify Relevant Documentation
  \item Characterize Restrictions on Use Cases
  \item Characterize Restrictions on Actors, Actions, Entities
  \item Develop legal memorandum
  \end{enumerate}
}

  &

  2. Domain Analysis:
  B. Coding Actions and Conditions
  \textnormal{
  \begin{enumerate}[label=\alph*.]
  \item Map rules permitting and restricting action to law-specific characteristics
  \item Map law-specific characteristics to license text and affirmations
  \item Map law-specific characteristics to general properties
  \end{enumerate}
}\\
\end{tabular}
\caption{Overview of the stages of the formalization process.}\label{formprocess}
\end{table*}

The rest of the article is structured as follows.
In Section~\ref{sec:workflow} we describe the workflow of a research data repository and how that workflow must account for restrictions from various domains. In Section~\ref{sec:systemworkflow} we describe our system, and illustrate how we encode the relevant provisions of a domain as a logic program, and how we use the logic program and facts provided by the user to help automate the workflow of a data repository. We describe in Section~\ref{sec:formalization} the methodology by which we formalize the relevant aspects of a domain, including development of legal memoranda based on use cases relevant to the domain, development of logic rules that capture the restrictions identified by the memoranda, and drafting of relevant license terms to be included in automatically generated license agreements. Section~\ref{sec:formal-ferpa} illustrates this formalization process for FERPA. We discuss domain composition in Section~\ref{sec:composition} and possible extensions to our system in Section~\ref{sec:extensions}. Section~\ref{sec:related} describes related work.






\section{Automating Repository Actions
}\label{sec:workflow}

The system we propose is designed to be integrated into a data repository workflow. This workflow can be conceptualized as having distinct stages, including deposit, acceptance, transformation, retention, and dissemination. In this section, we describe how actions at each of these stages (such as \emph{deposit}, \emph{accept}, \emph{transform}, and \emph{release}) can be automated using formal rules and license terms. In Section~\ref{sec:systemworkflow} we describe our system, which is designed to support the workflow we describe here.

\subsection{Deposit}

When a user, such as the PI from the hypothetical above, initiates deposit of a dataset through a repository's web interface, she will first encounter an interview. The interview presents the user with a series of questions that are designed to elicit facts about the dataset, the personal information it contains, the individuals in the data, the researchers and institutions that participated in the data collection and subsequent data transfers, and any privacy-related transformations that have been applied to the dataset, among many other related characteristics of the dataset and actors involved. The questions are written to be understandable by a lay user who has neither legal or technical expertise nor familiarity with terminology in this area. Questions are written in plain language, with minimal use of terms of art and are accompanied by detailed definitions, non-technical explanations, and real-world examples.

The interview questions are designed to determine which laws and best practices are applicable to the information in the dataset. For example, returning to the hypothetical above, the interview will ask questions to determine whether FERPA applies to the dataset. One such question is designed to elicit a fact about the provenance of the dataset, specifically whether any information in the dataset originated from records maintained by an educational agency or institution, as defined by FERPA. In the Appendix (Figure~\ref{FERPAquestion}), we show (a) a plainly-worded question, tailored to non-experts acting within research data transfer use cases, that was constructed based on (b) language from the FERPA regulations that is likely to be unfamiliar and difficult for a non-expert to parse.

A depositor uses what she knows about the attributes of a particular dataset and the relevant actors to provide responses to the questions presented by the repository. Based on the depositor's answers, the repository is able to conduct an analysis. This analysis identifies the key conditions that likely apply to the data and imply legal and best practice requirements. For example, an affirmative response to the sample question in Figure~\ref{FERPAquestion}, in combination with affirmative answers to other FERPA-related questions, would likely imply that conditions required by FERPA apply and constrain actions on the data.

When the choice of license relies on a particular legal fact that is supplied by the depositor, it is best practice to have that fact affirmed directly in the license agreement. This can be accomplished in one of two ways -- by ensuring that the license conditions for that use case require a separate affirmation license term, or by including the question and the user response verbatim. We refer to questions that are used to directly affirm  legal facts as affirmations -- and the license template is used to automatically insert any active affirmations into the final license.   

For example in this case, our license must include an affirmation that The deposit process will also require the depositor to affirm the truth of the facts that were inferred to be true on the basis of her responses to the questions. These facts are presented to the user as affirmations, such as ``I affirm that the data contain information derived from records maintained by an educational agency or institution that receives funds from the US Department of Education,'' and the users must confirm their applicability. Depending on the conditions that are identified and confirmed by the user, one of three outcomes will occur: (1) deposit of the dataset will be permitted, and data deposit and sharing licenses will be automatically generated based on the conditions that are required, (2) the dataset will be flagged for human review, or (3) the deposit request will be rejected.

\subsection{Acceptance}

Upon determining that the user is permitted to deposit the dataset and the repository is permitted to accept it, the repository will present the user with a bundle of licenses and related documentation. This bundle contains a complete license for deposit, using standardized terms, that establishes a contract between the depositor and the repository.\footnote{One may note that Deposit, Acceptance, and Dissemination could be further decomposed into transmission and receiving actions. Deposit/acceptance can be stated as the depositor transmitting a dataset to a repository, which receives it. And dissemination could be stated as the repository transmitting the dataset to a third party. We choose to use the higher-level semantic, as it 
  reduces the complexity of rules.
}

The bundle also includes a list of affirmations confirming the key conditions identified as applying to the data, as well as provenance documentation, which provides a record tracing the interview responses to the conditions and the conditions to the terms included in the license. The repository also shares documentation of its compliance with the laws identified as relevant. 
The principal investigator can, in turn, hand over the license and documentation for review by her home university's institutional review board, office of sponsored programs, or general counsel's office. This enhances accountability, as the process of producing and recording this documentation creates evidence of the parties' compliance with legal obligations and demonstrates their trustworthiness. Upon review and approval by the relevant offices at her home institution, the principal investigator can return to the repository, agree to the license, and deposit the data. At this time, human-readable and machine-actionable metadata reflecting the applicable requirements are then associated with the deposited dataset. These metadata are generated by the logical reasoning system described above, encompassing the permissible actions that may be taken with respect to the dataset and under what conditions and representing a data handling policy that is custom to the dataset.

\subsection{Retention}

Throughout the repository's long-term retention of the dataset, these metadata serve as the basis for making automated decisions regarding how the dataset will be stored, encompassing the steps the repository must take to comply with various data security, retention, destruction, and breach-reporting requirements in accordance with applicable laws and best practice. Documenting these requirements as metadata that will follow the dataset throughout its lifecycle enables the repository to make representations to the depositor that the dataset will be handled appropriately over the long term, and ensures accountability for each of the parties involved.

\subsection{Release and Transformation}

Based on release conditions that apply and are recorded in the metadata, the repository will be able to automate some of its decisions to release the dataset to users for secondary research purposes. For example, the metadata may reflect that the dataset can be transferred under an automatically generated license to any user, or only to certain, vetted users. In other cases, where it is not appropriate to automatically make the deposited dataset available for download, the repository may instead allow access to the data through a protected analysis mechanism (e.g., \citep{PSI}), or to a version of the dataset that has been produced by applying privacy-preserving transformations.




\begin{figure*}[t]
\centering
\fbox{\includegraphics[width=0.8\linewidth]{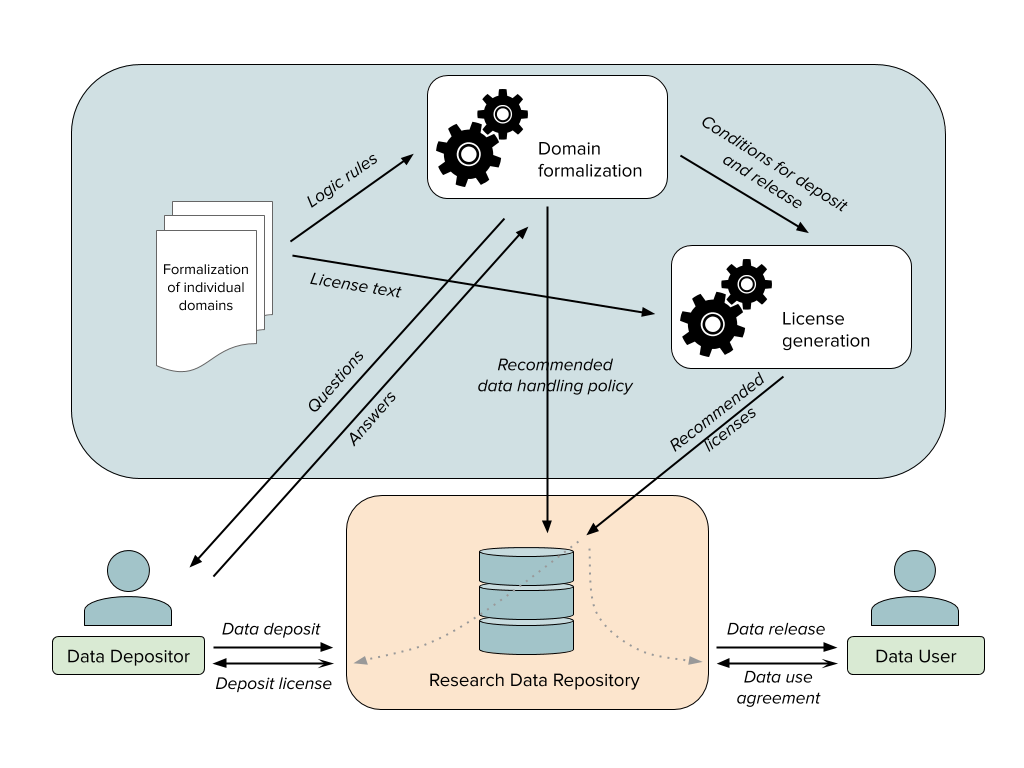}}
\caption{Conceptual system diagram.\label{workflow}}
\end{figure*}

\section{System Design}\label{sec:systemworkflow}

Figure~\ref{workflow} presents an overview of our system design. There are two key components: (1) a formalization of relevant aspects of domains and (2) an automated license generator.

We use a logic programming language (currently, Prolog) to formalize privacy-relevant domains, which enables us to state rules declaratively. Monotonicity of the logic facilitates modularity and compositionality. Also, logic programming enables us to use the formalizations in different modes, for example, to determine whether a given action is permitted or denied for a dataset, or to search for conditions under which a given action would be denied.

\subsection{Actions and Permissions}\label{sec:actions-permissions}

The key concepts in the formalization are \emph{actions} and \emph{permissions}. Actions include the deposit of a dataset in a repository and the release of a dataset to a data user. These two actions are represented respectively by \lstinline{deposit(DD, DS, R, CS)} and  \lstinline{release(R, DS, DU, DD, CS)}, where \lstinline{DD} represents the data depositor, \lstinline{DS} the data set, \lstinline{R} the repository, \lstinline{DU} the data user, and \lstinline{CS} a set of licenses and other conditions that restrict an action. The domain might indicate whether an action is permitted or denied. For example, \lstinline{permitted(ferpa, release(R, DS, DU, DD, CS))} represents that (our formalization of) the FERPA legislation permits the action \lstinline{release(R, DS, DU, DD, CS)}. Similarly, \lstinline{denied(ferpa, A)} represents that FERPA legislation denies action \lstinline{A}. A given domain may neither permit nor deny a given action, if, for example, the dataset does not fall under the scope of the domain.

\begin{figure}
\begin{lstlisting}[frame=single]
permitted(ferpa, release(_R, DS, _DU,
    _DD, CS), N) :- 
    bounded(CS, N), 
    ferpa_datasetInScope(DS), 
    \+(ferpa_identifiable(DS)). 
\end{lstlisting}
\caption{Sample FERPA rule permitting release of a dataset that does not contain identifiable information. The parameter \lstinline.N. and predicate \lstinline.bounded(CS, N). are used to ensure that search for conditions \lstinline.CS. is bounded.}\label{identifiablerule}
\end{figure}

We model domains with inference rules in the logic programming language. Each domain that we model is modularly expressed by its own set of inference rules, following the modeling process described below in Section~\ref{sec:formalization}. In this paper, we provide sample rules used to model FERPA. The rules describe the conditions under which FERPA permits the release of a dataset that is in the scope of FERPA (i.e., contains information about educational records of  a US educational agency or institution). For example, Figure~\ref{identifiablerule} provides a sample FERPA rule permitting the release of a dataset if it does not contain identifiable information, where ``identifiable information'' is defined by FERPA. Figure~\ref{auditstudiesrule} provides another FERPA rule permitting release if the dataset is released under an exception for research studies or auditing.\footnote{There are several other conditions under which FERPA permits release -- including for specific consented purposes. For an exhaustive list, see the online appendix.}

\subsection{Modeling Release Rules}\label{sec:model-release-rules}

Let's consider one of the rules in more detail: the rule for release under the studies exception or the audit exception to FERPA, as provided in Figure~\ref{auditstudiesrule}. The rule states that FERPA permits a release of dataset \lstinline{DS} under conditions \lstinline{CS} if the dataset is in the scope of the FERPA legislation (\lstinline[breaklines=true]{ferpa_datasetInScope(DS)}), the dataset contains identifiable information as defined by FERPA (\lstinline[breaklines=true]{ferpa_identifiable(DS)}), the dataset is shared under either the studies exception or the audit exception (\lstinline[breaklines=true]{ferpa_studiesException(DS); ferpa_auditException(DS)}, where \lstinline[breaklines=true]{;} indicates disjunction in Prolog), and the dataset is released either with a license that satisfies the studies exception or a license that satisfies the audit exception (\lstinline[breaklines=true]{ferpa_license_studiesException(CS); ferpa_license_auditException(CS)}). Note that the rule deliberately permits the dataset to be, for example, shared initially under the audit exception, but re-disclosed under either the studies or audit exception. This is a result of the legal analysis (Section~\ref{sec:ferpa-analyze}), which permits disclosure under either exception regardless of how the data was originally shared. 






\begin{figure}
\begin{lstlisting}[frame=single]
permitted(ferpa, release(_R, DS, _DU,
   _DD, CS), N) :- 
   bounded(CS, N), 
   ferpa_datasetInScope(DS), 
   ferpa_identifiable(DS), 
   (ferpa_studiesException(DS);
      ferpa_auditException(DS)), 
   (ferpa_license_studiesException(CS);  
      ferpa_license_auditException(CS)). 
\end{lstlisting}
\caption{Sample FERPA rule for studies exception and audit exception.}\label{auditstudiesrule}
\end{figure}

Some predicates used in these rules are themselves defined by additional rules. For example,
predicate \lstinline[breaklines=true]{ferpa_license_studiesException(CS)} (defined in Figure~\ref{FERPArules}, in the Appendix) holds if conditions \lstinline{CS} contain licenses that satisfy a number of requirements, including that the license lists the duration of the study for which the data is being released (\lstinline[breaklines=true]{conditionsRequire(CS, ferpa_license_duration)}), and that the license contains terms related to the destruction of data upon the completion of the study (\lstinline[breaklines=true]{conditionsRequire(CS, general_license_dataDestruction)}).

Other predicates represent facts about the world that cannot be inferred by our formalization, such as whether a dataset contains identifiable information (e.g., the predicate \lstinline[breaklines=true]{ferpa_identifiable(DS)} indicates that dataset \lstinline{DS} contains identifiable information, as defined by the FERPA legislation). These facts must be provided by a user familiar with the dataset, via the automated interview with the data depositor. We express this in our system design diagram (Figure~\ref{workflow}) by the interaction between the data depositor and our system.

Using a logic program to express the formalization of privacy-relevant aspects of domains enables us to explore privacy requirements in various ways. Most straightforward is to provide facts regarding a specific dataset, and then to use the logic program to answer which actions on that dataset are permitted and denied by various domains. These permitted and denied actions can form the basis for how the data repository chooses to handle the dataset, in terms of access control decisions and encryption on data at rest and in transit, among other choices.

Our logic program formalization can also be used to search for conditions under which (our formalization of) a domain is contradictory (i.e., there exists a dataset and an action on the dataset such that the domain both permits and denies that action) and for conditions under which the domain is silent (i.e., there exists a dataset and an action that should be in the domain's scope, but the domain neither permits nor denies that action).

\subsection{Modeling Transformations}\label{sec:transformations}

Our system can also express rules that capture reasoning about transformations of datasets, and how those transformations interact with a domain's privacy-relevant requirements.
In general terms, we use {\em transformation} to describe the output yielded by a particular set of computations applied to a dataset. How data are transformed often affects whether data are permitted to be stored or released. Many laws explicitly refer to transformations such as redaction, de-identification or anonymization, and encryption.
Formally, we reason over transformation through encoding that:
\begin{enumerate}
\item A dataset \textit{derives} from another dataset;
\item The derivation occurs through the use of a tool, with a given set of parameters;
\item The repository owner \textit{affirms} that the tool satisfies a transformation condition; and
\item Some action is \textit{permitted} on an output based on the transformation condition.
\end{enumerate}

Returning to the example of FERPA, consider that FERPA permits the release of de-identified information. There may be different transformations under which a dataset would be permitted to be released by FERPA. Because the text of FERPA and related interpretive guidance does not provide a single technique or list of techniques that is deemed to be sufficient to meet the de-identification standard, such a transformation condition does not come from the formalization of the law itself. Rather, the transformation condition may come from an institutional policy or best practice guideline. For example, some organizations may use aggregation or suppression techniques to meet FERPA's de-identification requirements as an institutional policy. Other organizations may choose to rely on best practice guidelines regarding state of the art techniques for protecting privacy. For instance, experts have demonstrated that the use of differential privacy (with a reasonable value for the privacy parameter \textit{epsilon}) is sufficient to satisfy FERPA's de-identification requirements~\cite{NissimBridging}.

To illustrate this, consider a rule permitting the release of FERPA-protected data that has been transformed using differential privacy, shown in Figure~\ref{DPrule}.
The rule states that FERPA permits the release of a dataset that is in the regulation's scope, if the dataset is derived from a dataset \verb.DS2. using a differentially private transformation where the total information release budget (i.e., over all data releases) is considered to be sufficient to meet the deidentification standard of FERPA.\footnote{This sample rule takes \lstinline{0.1} to be a reasonable value of the privacy budget parameter epsilon and therefore sufficient to meet the de-identification standard of FERPA, in accordance with recent best practice guidelines~\cite{NissimPrimer}. Note: the repository needs to be aware of the \emph{total} budget consumed over all releases, not just the budget consumed to derive dataset \texttt{DS} from dataset \texttt{DS2}.} Further rules for accepting and depositing data transformed using differential privacy can be similarly formalized. 

\begin{figure}
\begin{lstlisting}[frame=single]
ferpaSufficientEps(0.1).

permitted(ferpa, release(_R, DS, _DU,
          _DD, CS), N) :-
    bounded(CS, N), 
    ferpa_datasetInScope(DS), 
    derivedFrom(DS, DS2,
    	differentialPrivacy(Params)), 
    member([totalBudget, EPS], Params),
    ferpaSufficientEps(FE),    
    EPS <= FE.
\end{lstlisting}
\caption{Sample best practice rule permitting the release of a FERPA-protected dataset that has been transformed using differential privacy.}\label{DPrule}
\end{figure}

Determining that dataset \verb.DS. is derived from dataset \verb.DS2. with  parameters \verb.Params.
may require interview questions and affirmations from the user. Alternatively, and more flexibly, a repository that provides the ability to transform datasets could provide these facts. For example, a repository could use a differential privacy tool, such as PSI~\citep{PSI}, to transform the data and provide our system with appropriate facts about how one dataset was derived from another using this tool.
%
%
%
This can support not only the creation of public-use output from protected data, but could be used to enable the automated use of transformation tools to enable deposit or release of otherwise protected data.

\subsection{Modeling Purpose Restrictions}\label{sec:model-purpose}

Most domains impose restrictions on data that are conditioned on the purpose of the data controller's or data user's use of the data. For example, FERPA permits non-directory personally identifiable information to be used for any purpose to which the data subject has specifically consented; further, such information can be used without consent for certain purposes (and by certain entities). Under the FERPA studies exception, consent is not required for research studies conducted by certain parties that have the purpose of developing, validating, or administering predictive tests; administering student aid programs; or improving instruction.\footnote{\textit{See} 34 C.F.R. \textsection\ 99.31(a)(6).}

Generally, purpose restrictions such as these can be modeled in three (non-exclusive) ways: as an independent  condition; as an opaque text statement; or as a set of permitted and denied component purposes drawn from a common taxonomy.

First, where a specific set of purposes is defined by the domain itself, we may model that term as a unique condition. This approach is most appropriate when the meaning or language associated with that purpose is idiosyncratic to that domain, or is inextricably combined with other simultaneous conditions. For example, we model the studies exception above using the dataset predicate \lstinline{ferpa_studiesException(DS)}, and create a corresponding license term that satisfies the predicate. This is because the definition of a study is specific to FERPA, and because this restriction on purpose is never used independently of the other restrictions documented in the studies exception.
%
%

Second, where a purpose restriction is defined not by the law but by the data subject or data depositor, statements of permitted purposes and  denied purposes are recorded when the data set is ingested. Release can then be permitted under the condition that the recipient agrees to the  terms supplied. This restriction is reflected in the condition set for a release action in our formalization and realized as a specific license term added to the automatically-generated agreement.

%

Last, where purpose restrictions are generalizable we define them in terms of formal taxonomies. Although the details of taxonomy selection are beyond the scope of this work, we illustrate the purpose restrictions using the EuroVoc Thesaurus \cite{eurovoc}, which is a officially created, multi-lingual, formally defined (based on ISO-5964 and ISO-2788) thesaurus which describes human activities subject to EU regulation. 

Using this taxonomy we can characterize both purpose categories that are always permitted and purpose categories that are always denied for a given dataset. This can allow automation of some decisions about whether a particular purpose is suitable for a given dataset. However, since
the taxonomic categories are often broader than domain-defined or user-supplied conditions, some decisions will require human review based on the specific purposes text supplied (e.g. in the consent agreement under which data is deposited.)

For example, a group of subjects might consent to their identifiable data being used for education research, or for medical research specific to schizophrenia. Since the EuroVoc taxonomy does not does not specify schizophrenia but includes the category of mental health, at the time of ingestion, a human-readable purpose statement could be ``Must be used for the purposes of research in  education or on schizophrenia only'', the always-permitted purpose category be \texttt{3206.Education AND 2194.Research}, the always-denied purpose category be \texttt{NOT ((2194.Research) OR NOT (3206.Education OR 5878.Mental Health)}, and human review is required for medical research purposes in order to determine whether they are for research on schizophrenia. 




\subsection{Automated License Generation}\label{sec:license-generation}

A key component of our system is automated license generation. Various kinds of licenses are needed during the lifecycle of data in a data repository. For instance, a deposit license between the repository and the data depositor at the time a dataset is submitted to a repository, and a data use agreement between the repository and a data user is needed when a dataset is released to the data user. These licenses may be required by legislation, best practices, or repository policy to contain certain information or text. We use output of the logic program formalization to determine whether these license requirements apply to a given action on a dataset. Based on the license requirements, we select snippets of license terms to include in the license. For each domain we formalize, these snippets form part of the formalization, containing text and the conditions under which to include the text in a license. For illustration, an example of such a formalization for FERPA is presented below in Section~\ref{LicenseTerms}. We use a license template (which contains placeholders indicating where to include the appropriate snippets of license text) and some mild text processing (e.g., formatting, and connectives for terms) to produce an appropriate license.

\section{Formalizing Privacy Law Domains for Data Controllers}\label{sec:formalization}

\subsection{Constraining the Scope of Action}

The domain formalization process begins with referring back to the specific use cases that are within the scope of research data sharing. From there, the analysis proceeds by identifying the relevant legal documentation and requirements that apply to the scope of actions circumscribed by those use cases.

Taking the data deposit use case for instance, there are implied  \textit{actors} (e.g., the data depositor and the data repository), \textit{actions} (e.g., the data depositor's deposit of the data set and the data repository's acceptance of the data set), \textit{objects} (e.g., the data set and the records contained in the data set), and \textit{properties} (e.g., the data set contains records within the scope of FERPA).

\subsection{Analyzing Legal Restrictions Applicable to Use Cases}\label{legalanalysis}

The formalization is supported by detailed legal memoranda analyzing individual laws, institutional and contractual approaches to data sharing, and the relevant academic literature. To develop these memoranda, legal research is conducted to identify specific restrictions on use cases, actors, actions, and objects relevant to understanding the applicability of these regulations, policies, and contracts to researchers and data repositories. This involves, for example, analyzing federal and state regulations that restrict the access, use, and disclosure of sensitive data and the applicability of these regulations to researchers and data repositories. It also involves analyzing individual university data classification policies, institutional review board policies, and the academic literature to identify definitions, interpretations, and practices for implementing requirements of data privacy laws in research settings.

For each law analyzed, the output is a legal memorandum that summarizes research and analysis to apply the law to a set of facts based on the identified scope of action. This analysis seeks to determine which provisions of a specific law apply to the relevant researcher and research data repository use cases, to identify the conditions under which these provisions are applicable, and to identify the restrictions that must be imposed in cases in which such conditions are applicable.

\subsection{Expert Coding}\label{ExpertCoding}

The aforementioned legal memoranda serve as the basis for the formalization process. The formalization relies on expert coding to translate requirements from law, as expressed in the legal memoranda, together with requirements from policy and best practice, into formal rules, conditions, license terms, and affirmations. This process involves multiple steps:

\begin{enumerate}

\item Identification of a key legal requirement, policy requirement, best practice, or common practice,

\item Identification of the key dataset or data controller properties relevant to determining whether the legal requirements or practices apply to a given dataset,

\item Mapping rules permitting and restricting action to law-specific characteristics,

\item Mapping law-specific characteristics to general properties,

\item Construction of rules that link properties to actions,

\item Mapping law-specific characteristics to license text (including affirmations),

\item Construction of  questions (in addition to affirmations) to elicit properties from data controllers,

\item Construction of rules linking question responses to properties, and

\item Construction of rules linking properties to permitted and denied actions under appropriate license conditions.
\end{enumerate}

In the next section we illustrate the key points of the coding process using a concrete example. 

\section{Illustration: Formalizing FERPA}\label{sec:formal-ferpa}

To illustrate the formalization process, we detail the steps involved in our formalization of a selected law, the Family Educational Rights and Privacy Act (FERPA).\footnote{(20 U.S.C. \textsection\ 1232g; 34 C.F.R. Part 99).} Based on an analysis of a corpus of research data repository policies, FERPA was identified as a key regulatory requirement in the context of the research data repository use cases identified. FERPA protects personally identifiable information from education records maintained by educational agencies and institutions that receive funding from the U.S. Department of Education.\footnote{20 U.S.C. \textsection\ 1232g(a)(1)(A)).} Under FERPA, non-directory personally identifiable information\footnote{FERPA defines \textit{personally identifiable information} as follows: ``The term includes, but is not limited to-- (a) The student's name; (b) The name of the student's parent or other family members; (c) The address of the student or student's family; (d) A personal identifier, such as the student's social security number, student number, or biometric record; (e) Other indirect identifiers, such as the student's date of birth, place of birth, and mother's maiden name; (f) Other information that, alone or in combination, is linked or linkable to a specific student that would allow a reasonable person in the school community, who does not have personal knowledge of the relevant circumstances, to identify the student with reasonable certainty; or (g) Information requested by a person who the educational agency or institution reasonably believes knows the identity of the student to whom the education record relates. 34 C.F.R. \textsection\ 99.3.} from education records generally cannot be disclosed by an educational agency or institution without consent, unless an exception to the FERPA consent requirement applies.

\subsection{Analyzing the Use Case}\label{sec:ferpa-analyze}
Legal research and analysis summarized in the FERPA memorandum identified a number of ways in which personally identifiable information (PII) from education records protected by FERPA could be shared via a research data repository. Specifically, any of the following conditions, if applicable, could permit such a transfer:

\begin{itemize}
\item \textbf{Disclosure with consent.} FERPA permits the disclosure of PII with the consent of a student over the age of 18 (or the student's parent, if the student is under the age of 18). The consent must be in writing, signed, and dated, and it must specify (1) the records that may be disclosed; (2) the purpose of the disclosure; and (3) the party or class of parties to whom the disclosure may be made.\footnote{34 C.F.R. \textsection\ 99.30(a), (b).}
\item \textbf{Disclosure of de-identified information.} FERPA does not require consent for the disclosure of de-identified information from which all PII has been removed,\footnote{34 C.F.R. \textsection\ 99.31(b).}  ``provided that the educational agency or institution or other party has made a reasonable determination that a student's identity is not personally identifiable, whether through single or multiple releases, and taking into account other reasonably available information.''\footnote{34 C.F.R. \textsection\ 99.31(b)(1).}
\item \textbf{Disclosure of directory information.} FERPA does not require consent prior to the disclosure of directory information,\footnote{Directory information is information from student records that can be released to the public because it would not generally be considered harmful or an invasion of privacy if disclosed (34 C.F.R. \textsection\ 99.3). Directory information includes information generally considered to be directly identifying, such as names, addresses, telephone numbers, and photographs, as well as information that may be indirectly identifying, including dates and places of birth, major fields of study, and honors and awards.} as long as the educational agency or institution has (1) given parents and students public notice of the types of PII that the agency has designated as directory information and (2) given the parent or eligible student the opportunity to opt out of the disclosure or publication of their directory information.\footnote{34 C.F.R. \textsection\ 99.37(a).}
\item \textbf{Disclosure for audit or evaluation.} Consent is not required if the disclosure is to the Comptroller General of the United States, the Attorney General of the United States, the Secretary, or state and local educational authorities for audit or evaluation purposes.\footnote{34 C.F.R. \textsection\ 99.31(a)(3).}
\item \textbf{Disclosure for studies.} Consent is also not required if the disclosure is to organizations conducting studies for or on behalf of educational agencies or institutions to: develop, validate, or administer predictive tests; administer student aid programs; or improve instruction.\footnote{34 C.F.R. \textsection\ 99.31(a)(6)(i).}
\item \textbf{Disclosure to school officials.} PII may be disclosed to school officials with a legitimate educational interest in the records. A third party may be considered a school official if the party (1) performs an institutional function for which the agency or institution would otherwise use employees; (2) is under the direct control of the agency or institution with respect to the use and maintenance of education records; and (3) is subject to FERPA's requirements governing the use and redisclosure of PII from education records.\footnote{34 C.F.R. \textsection\ 99.31(a)(1)(i)(B).} In particular,  de-identification may be outsourced to a vendor serving as a school official in accordance with these requirements.\footnote{\textit{See} 73 Fed. Reg. 74806, 74834 (Dec. 9, 2008).}
\end{itemize}

Each of the relevant provisions of FERPA carries conditions that are formalized using the expert coding process detailed in Section~\ref{ExpertCoding} above. Further, in Section~\ref{LicenseTerms}, we discuss how these requirements are formalized as specific license terms to be incorporated in custom data sharing licenses.

In the following sections, we discuss two aspects of this expert coding process, describing how we relied on the legal memorandum to draft sample domain-specific rules (Section~\ref{DomainRules}) and sample license terms (Section~\ref{LicenseTerms}) through a series of examples.

\subsection{Drafting Domain-Specific Rules}\label{DomainRules}

A selection of formal logic rules encoding FERPA legal requirements is illustrated in Figure~\ref{FERPArules}.\footnote{The full set of rules is available in our replication appendix online. Each rule is accompanied by a human-readable rationale, extracted from the legal memo, and by citations to the relevant legal documentation.} In the remainder of this section we illustrate the process we used to create these rules.

We first engaged a small team of law students, supervised  by a qualified lawyer to identify the documents most relevant to understanding FERPA legal requirements. These documents include the enabling sections of the United States code, section 34.99 of the Code of Federal Regulations, and critical guidance issued by the Department of Education's Privacy Technical Assistance Center.  Based on these documents, the legal team created a memorandum that described the legal restrictions that FERPA places on depositing, accepting, retaining, and disseminating data.

This memo characterized a number of properties that are critical to determining whether deposit of data (etc.) is permitted or denied under FERPA. Each of these properties is then encoded in the  the logic program for the FERPA domain. For example, the property of a dataset \verb.DS. containing identifiable data under FERPA standards is encoded as the predicate \lstinline[breaklines=true]{ferpa_identifiable(DS)}. Other properties such as data collected with prior consent for release and data  being received under the FERPA studies exception, are similarly encoded.

Once key properties have been encoded, the team characterizes the constraints described in the memo by referring to the relevant action, properties, and license conditions (if any). For example, Figure~\ref{auditstudiesrule} shows a rule (discussed in Section~\ref{sec:model-release-rules}) that encodes the permitted release of a dataset under either the audit exception or the studies exception.


\subsection{Drafting Relevant License Terms}\label{LicenseTerms}
 
The legal restrictions identified in the legal memoranda are encoded in modular license terms that are incorporated in the custom data sharing licenses produced by the system's automated license generator. 

 Separate license templates are associated with the deposit, accept and release actions. And the terms included in these templates are designed to be agreed to by separate actors: the depositor, repository, or data recipient actor (respectively). These agreements are generally asynchronous, and may be presented in different forms. For example, a repository may choose to include terms it satisfies as part of its website terms of service, rather than presenting them at each acceptance. 

Consider, for example, a data release under the studies exception to FERPA: an educational agency or institution may disclose PII, without consent, to organizations conducting studies for or on behalf of educational agencies or institutions to: develop, validate, or administer predictive tests; administer student aid programs; or improve instruction.\footnote{34 C.F.R. \textsection\ 99.31(a)(6)(i).}

This restriction on the use of PII is encoded as the following sample license term to be included in data sharing licenses for sharing data in accordance with the studies exception to FERPA:

\begin{licensetext}
    Personally identifiable information
(FERPA) may only be used to conduct a
study for, or on behalf of, the
educational agency or institution that
originally maintained the information.
The study must be for the purpose of
developing, validating, or administering
predictive tests; administering student
aid programs; or improving instruction.
  \end{licensetext}

In addition, to fall within the scope of the studies exception, an educational agency or institution must enter into a written agreement with the recipient of the data. FERPA requires that this written agreement (1) specify the scope and purpose of the study as well as the information to be disclosed; (2) require the organization to limit the use of the PII to the purposes in the agreement; (3) ensure that the study must be performed in a way that does not allow personal identification of parents and students to anyone other that representations of the organization that have legitimate interests in the information; (4) require all PII to be destroyed when it is no longer needed and specify the time period in which it must be destroyed.\footnote{34 C.F.R. \textsection\ 99.31(a)(6)(iii)(C).}

Based on the legal memorandum relying on the text of the regulations as well as subsequent guidance from the Department of Education,\footnote{U.S. Department of Education, The Family Educational Rights and Privacy Act: Guidance for Reasonable Methods and Written Agreements (August 2015).} we encoded these requirements for written agreements into the following sample license terms. The first four license terms encode the first requirement for written agreements, and the last license term encodes the remaining requirements.

\begin{licensetext}
The Data Recipient agrees to use the
data only for the purpose of conducting
a study to
[dataUser:supplied:FERPA:studyPurpose].
\end{licensetext}

\begin{licensetext}
The Data Recipient agrees to use the
data only during the period of
[dataUser:supplied:FERPA:studyStartDate]
through
[dataUser:supplied:FERPA:studyEndDate].
\end{licensetext}

\begin{licensetext}
The Data Recipient agrees to use the
data only for studies on the following
topics:
[dataUser:supplied:FERPA:studyTopics].
\end{licensetext}
\vspace{-1ex}
\begin{licensetext}
The information disclosed is as follows:
[dataOwner:supplied:dataDescription].
\end{licensetext}

\vspace{-1ex}

\begin{licensetext}
The Data Recipient will:

(1) limit access to personally
identifiable information in the Data to
individuals with legitimate interests,

(2) conduct the study in a manner that
does not permit the personal
identification of parents and students
by anyone other than representatives
of the Data Recipient with legitimate
interests, and

(3) take steps to maintain the
confidentiality of the personally
identifiable information in the Data
at all stages of the study, including
within the final report, by using
appropriate disclosure avoidance
techniques.
\end{licensetext}


Finally, each term is assigned a category tag (and optional grouping id) to indicate where it should be included in the license template corresponding to that action (e.g. deposit, release, acceptance). For example, the first text above is categorized as a ``TERMS:PERMITTED USES'', and would be inserted (with other terms in the same category) in the following portion of the release license template. 
\begin{licensetext}
Permitted uses. 

~

Data Recipient is granted a nonexclusive, revocable license,
to have and use the Data provided the Data Recipient,
shall comply with all of the terms and conditions of this License.

~

[TERMS:PERMITTED USES]
\end{licensetext}

\section{Combining and Customizing Domains}\label{sec:composition}

Our system is designed to be modular, allowing the composition of domains (i.e., of different legislation, regulations, best practices, local policies, etc.). A data repository may choose to compose modules for legislation relevant to the geographic location and focus area of the repository, as well as policies for the institution. For example, a data repository at \ifanonymous a university in Massachusetts \else Harvard University \fi may use modules related to Massachusetts and US privacy legislation, as well as a module that captures \ifanonymous that university's \else Harvard's \fi data handling requirements.  We briefly describe what constitutes a domain module and how they can be composed.


A module includes: logic programming rules defining the conditions under which actions are permitted or denied; interview questions (and how they relate to facts about data sets); and license terms. For domain \lstinline{D}, the logic programming rules define predicate \lstinline{inScope(D, A)} that indicates whether action \lstinline{A} is in the scope of domain \lstinline{D}. For example, whether \lstinline{inScope(ferpa, A)} holds will depend on whether the dataset associated with action \lstinline{A} contains any information derived from records maintained by an educational agency or institution, etc. A given action may be in the scope of multiple domains, and each domain may indicate whether it permits or denies the action. A reasonable way to compose permissions would be for a repository to permit action \lstinline{A} only if all domains for which the action is in scope permit it, and to deny the action if there is at least one domain for which \lstinline{A} is in scope that denies the action. However, repositories are free to compose permissions however they choose. In particular, a repository must decide how to handle actions which are not in the scope of any domain, or if there is a domain for which the action is in scope but is neither permitted nor denied. We show an example of a top-level module that composes other modules in the Appendix (Figure~\ref{toplevelmodule}).

License terms and interview questions have unique identifiers, and a module may refer to license terms and questions from another module. Similarly, a module may refer to predicates defined by another module. We do not currently have a mechanism for a module to modify the license terms, interview questions, or logic rules defined by another module.



A repository may also want to add rules that indicate acceptability of tools for transformations (see Section~\ref{sec:transformations}), as this is likely driven by local institutional policy than legal formalization. 



\section{Extensions and Applications}\label{sec:extensions}

The tool we describe is in a functional prototype stage. It is publicly available.\footnote{\ifanonymous GitHub URL removed for anonymous submission.\else\url{https://github.com/MIT-Informatics/LegalTags}\fi}
We have not yet settled on a standardized format for a module's interview questions, but have implemented a questionnaire for FERPA and Massachusetts privacy law, as well as logic programming rules and license terms for these two domains.
The logic programming rules model permissions for deposit, acceptance, and release actions (Sections~\ref{sec:actions-permissions} and \ref{sec:model-release-rules}) and some transformations (\ref{sec:transformations}). We have an expert-system tool that allows exploring the implications of the logic programming model and a tool for automated license generation (Section~\ref{sec:license-generation}), both implemented in Java. We have not implemented a taxonomic approach for purpose specification (Section~\ref{sec:model-purpose}).
In this section we discuss potential extensions and applications. 

\subsection{Deployment}

The prototype framework can be used to generate complete licenses, and to evaluate specific decisions. However the software framework is not tested for operation in a production environment, nor provides a user-friendly interface.  A potential avenue  for  deployment is integration with the Policy Models online interviewing system \cite{BarSinaiSC16} and the Dataverse repository  \cite{dataverse}. Using this approach, our system would generate and/or validate interview instruments that would be deployed with  the Policy Models server; interview results would be used by our system to generate  licenses which would then be transmitted to and stored in the Dataverse repository. 

\subsection{Local Review of Modules}

 Before any module is used, repository owners may wish to have local legal experts review the module rules. We aim to facilitate this in future by providing a prepared corpus of fact-patterns, and corresponding licenses resulting from these, in addition to the annotated decision rules and terms. This corpus, and external expert evaluations of it, will be available for inspection by users of the system and will aid local review.

\subsection{Adding Domains of Law and Practice}

The modular design of the system enables it to be readily extended to new domains, without any changes to the software or design of the system. Each domain is encoded in a separate module, containing inputs, rules and conditions. We are currently developing modules for data protected by HIPAA, the Common Rule, and Massachusetts privacy law. 
Modules for other laws and best practices  may be created, disseminated, and deployed, independently of our framework.

\subsection{Adding Transformation and Purpose Restrictions}

As described above, the system is capable of reasoning over the deposit and release of data that has undergone a protective transformation, such as Differential Privacy, or which is subject to purpose restrictions, such as consented uses. Many purposes and transformation are applicable across multiple domains. A useful extension of the legal analysis would be to create detailed rules across a set of common transformations (including encryption, deidentification, and k-anonymity) and a range of standardized purposes (e.g. ``non-commercial'', ``statistical'').

\subsection{Towards a Privacy Commons License}

As modules are added, the complexity of the interviews and licenses generated increase because each domain may require its own  set of affirmations and terms.
Difference in terminology used to express similar concepts across multiple domains can lead to  interviews that are unnecessarily long.

To avoid this tiresome user burden, common terminology could be developed that applies to multiple domains, and independent  modules could encode the best practices for applying this ``privacy commons''. 

While the   legal analysis required to develop a full set of commons licenses is substantial, the formalization for applying common terms is simple. A module can express common terms by creating a rule that asserts that affirming Common Term X satisfies some domain specific property, as discussed in the Combining Modules section. Common rules can be introduced  and reviewed incrementally---and legal use cases tested empirically using the system to evaluate whether they can be satisfied solely by common terms.


\section{Related Work}\label{sec:related}

Logic and logic programs have long been used to formalize
legislation. An early example is a Prolog formalization of the British
Nationality Act in the 1980s \citep{SergotSKKHC1986}.

\citet{Leith1986} critiques such attempts, arguing that the idea of a
``clear rule of law'' is invalid and due to factors outside the
legislation---including the legal process---it is not possible provide
legal expert systems that can predict ``real judicial outcomes.'' We
agree that it is futile to attempt to fully formalize legislation in
the hope of replacing lawyers and the judiciary with a computer
system, and instead aim to formalize a very narrow aspect of
privacy-relevant legislation and best practice: providing guidance for
how a data repository should decide to accept and share data sets.

Indeed, our narrow focus allows us to use a straightforward logic
(i.e., the fragment expressible in Prolog) to express the
privacy-relevant requirements. By contrast, a relatively large body of
work uses more sophisticated logics to express more aspects of privacy
legislation.

\citet{BarthDMS07,BarthDMN07} use an alternating-time temporal logic
to express privacy and utility goals. Privacy is expressed as linear
temporal logic (LTL) formulas, and utility requirements are expressed
as agents having appropriate strategies to accomplish certain
outcomes. They show that relevant questions of privacy policies (e.g.,
policy consistency, compliance, composition, and refinement) reduce to
well-studied LTL problems. Later work \citep{ChowdhuryGNRBDJW2013}
extends some of the notions of compliance to a more expressive logic.

\citet{DeYoungGJKD2010} extend the work of Barth et al. to formalize
portions of HIPAA and the Gramm-Leach-Bliley Act (GLBA) that are relevant to the
transmission of information. To do so, the propose a custom logic
PrivacyLFP, a least fixed point logic with a trace-based semantic
model, which allows PrivacyLFP to express temporal properties. They
require the power of a fixed point logic because the legislation is
self-referential.

\citet{AucherBT2011} take a different approach to a logical
formalization of privacy, using an epistemic deontic dynamic
logic. That is, their logic reasons about knowledge of agents,
obligations on who knows what, and accounts for updates of knowledge
due to actions, such as transmitting information.

The scope of the work of \citeauthor{DeYoungGJKD2010} and
\citeauthor{AucherBT2011} is much greater than our framework, which
focuses on deciding whether a data repository should perform certain
limited actions. As such, we are able to use a much simpler logic in
our formalization.

\citet{DattaBCDGJKS2011} present a formal framework to characterize
the correctness of auditing of privacy policies (expressed with
PrivacyLFP).  \citet{ChowdhuryJGD2014} considering the runtime
monitoring of temporal logic policies, using privacy policies as case
studies. We do not focus on monitoring or auditing
policy compliance.

\citet{BackesBHM2015} present a framework for reasoning about privacy
case law. The framework takes into account precedence and court
hierarchy, and facilitates reasoning about which court precedents
apply to a particular case. In our work, we are not concerned with
reasoning directly about case law and precedence, but rather focus on
encoding current practices and understanding in our formalization of
the conditions under which actions are permitted or defined by
legislation.

\citet{MasseyRAS2014,MasseyOA2015} consider extracting software
engineering requirements from legislation and regulations. They
determine that typical graduate-level software engineering students
can not write legally compliant software with any confidence and
develop a taxonomy of ambiguities that might complicate the extraction
of requirements. The taxonomy does not aim to provide a methodology
for formalization or automated reasoning about legislation or
regulations.

Other frameworks focus on the enforcement of privacy policies in systems.
\citet{SenGDRTW2014} develop a framework that allows the specification
of privacy policies (in terms of restrictions on user data) and the enforcement of these policies in Map-Reduce-like
big-data settings.


\section{Summary}

Describing existing law using formal logic is challenging. We address this challenge by limiting the domain of action in two ways (1) we focus on a specific domain of use cases, actors and actions --  the data repository; (2) and  focus on subset of conditions that can be automatically permitted or denied -- while explicitly supporting the possibility of escalation to ``human'' computation in cases that are not potentially allowable but cannot be determined automatically.

We have developed a prototype system that is capable of automating most repository decisions controlled by FERPA,  and of  generating appropriate licenses where use is permitted. This system can be extended to other laws and practices, and to reason over data protections such as differential privacy.

\bibliographystyle{plainnat}
\bibliography{privlaw}

\begin{figure*}
  \noindent\addtolength{\fboxsep}{3pt}\centering
  \framebox{
\begin{minipage}{0.96\linewidth}\begin{multicols}{2}
{\em (a) Sample interview question:}\medskip

Do the data contain any information derived from records maintained by an educational agency or institution, or by a person or entity acting for the agency or institution, that receives funds from the US Department of Education?\bigskip

\underline{Terms}\medskip

\textbf{Educational agency or institution}\\
An educational agency or institution is a public or private agency or institution that provides educational services or instruction, or both, to students. This includes, but is not limited to, a primary or secondary school, college or university, school district, or state department of education.\medskip

\textbf{Funds from the US Department of Education}\\
This means funds provided for any purpose, research or otherwise, by the US Department of Education to the educational agency or institution. It includes funds provided by grant, cooperative agreement, contract, subgrant, or subcontract; or funds provided to students attending the agency or institution. The funds may be paid to the institution by those students for educational purposes, such as under the Pell Grant Program and the Guaranteed Student Loan Program. This includes all public schools and a majority of private institutions, though most private and parochial schools at the elementary and secondary levels do not receive such funding.
\\

{\em (b) Corresponding language from the regulations:}\medskip

(a) Except as otherwise noted in § 99.10, this part applies to an educational agency or institution to which funds have been made available under any program administered by the Secretary, if-\smallskip

\begin{myindentpar}{4ex}

(1) The educational institution provides educational services or instruction, or both, to students; or\smallskip

(2) The educational agency is authorized to direct and control public elementary or secondary, or postsecondary educational institutions.\smallskip

\end{myindentpar}

(b) This part does not apply to an educational agency or institution solely because students attending that agency or institution receive nonmonetary benefits under a program referenced in paragraph (a) of this section, if no funds under that program are made available to the agency or institution.\smallskip

(c) The Secretary considers funds to be made available to an educational agency or institution if funds under one or more of the programs referenced in paragraph (a) of this section-\smallskip

\begin{myindentpar}{4ex}

(1) Are provided to the agency or institution by grant, cooperative agreement, contract, subgrant, or subcontract; or\smallskip

(2) Are provided to students attending the agency or institution and the funds may be paid to the agency or institution by those students for educational purposes, such as under the Pell Grant Program and the Guaranteed Student Loan Program (Titles IV-A-l and IV-B, respectively, of the Higher Education Act of 1965, as amended).\smallskip

\end{myindentpar}

(d) If an educational agency or institution receives funds under one or more of the programs covered by this section, the regulations in this part apply to the recipient as a whole, including each of its components (such as a department within a university).

\end{multicols}\end{minipage}}

\caption{Sample interview question for FERPA and corresponding language from the regulations (34 C.F.R. § 99.1).\label{FERPAquestion}}
\end{figure*}

\begin{figure*}
\begin{lstlisting}[frame=single]
permitted(ferpa, release(_R, DS, _DU, _DD, CS), N) :- 
    bounded(CS, N), 
    ferpa_datasetInScope(DS), 
    \+(ferpa_identifiable(DS)). 

permitted(ferpa, release(_R, DS, _DU, _DD, CS), N) :- 
    bounded(CS, N), 
    ferpa_datasetInScope(DS), 
    ferpa_identifiable(DS), 
    (ferpa_studiesException(DS); ferpa_auditException(DS)), 
    (ferpa_license_studiesException(CS);  
           ferpa_license_auditException(CS)). 

permitted(ferpa, release(_R, DS, _DU, _DD, CS), N) :- 
    bounded(CS, N), 
    ferpa_datasetInScope(DS), 
    ferpa_identifiable(DS), 
    ferpa_allConsented(DS), 
    ferpa_license_IRB(CS). 

permitted(ferpa, release(_R, DS, _DU, _DD, CS), N) :-
    bounded(CS, N), 
    ferpa_datasetInScope(DS), 
    derivedFrom(DS, _, differentialPrivacy(Params)), 
    member([totalBudget, EPS], Params),
    ferpaSufficientEps(FE),    
    EPS <= FE.

denied(ferpa, release(_R, DS, _DU, _DD, CS), N) :-
    bounded(CS, N),
    ferpa_datasetInScope(DS),
    ferpa_identifiable(DS),
    \+(ferpa_allConsented(DS)),
    \+(ferpa_studiesException(DS)),
    \+(ferpa_auditException(DS)).
    
ferpa_license_studiesException(CS) :- 
    conditionsRequire(CS, ferpa_license_notice), 
    conditionsRequire(CS, ferpa_license_purpose), 
    conditionsRequire(CS, ferpa_license_scope), 
    conditionsRequire(CS, ferpa_license_duration), 
    conditionsRequire(CS, ferpa_license_information), 
    conditionsRequire(CS, general_license_researchProposal), 
    conditionsRequire(CS, general_license_minimumPersonnel), 
    conditionsRequire(CS, general_license_minimumInformation), 
    conditionsRequire(CS, general_license_dataDestruction).    
\end{lstlisting}

\caption{Sample rules for FERPA.}\label{FERPArules}
\end{figure*}

\begin{figure*}
  \begin{lstlisting}[frame=single]
%% In this example top-level module, the policy is to comply with
%% the FERPA and Massachusetts privacy law (CMR) formalizations.
%% An action A is permitted by University X if every law that
%% is in scope permits it, and is denied by University X if it
%% is denied by either FERPA or CMR.
permitted(universityX, A, N) :-
    (permitted(ferpa, A, N); \+inScope(ferpa, A)),
    (permitted(cmr, A, N); \+inScope(cmr, A)).

denied(universityX, A, N) :-
    denied(ferpa, A, N); denied(cmr, A, N).


%% For the purposes of this example module, we will regard every
%% action as being in scope. This means that there are actions that
%% University X regards as in scope, but does not either permit nor
%% deny. Depending on the University X's policy, we may want to
%% deny any action that is not explicitly permitted. 
inScope(universityX, A).

%% Set the level of Sufficient budget for release under FERPA.
%% See the FERPA formalization.
ferpaSufficientEpsBudget(0.1).

%% Indicate that the PSI tool is a differentially private tool,
%% i.e., if we use the PSI tool to derive DS from DS2, then
%% we regard the deriviation as being differentially private.
derivedFrom(DS, DS2, differentialPrivacy(Params)), 
derivedFrom(DS, DS2, psiTool(Params)),     
  \end{lstlisting}
\caption{Example logic program for a top-level module that composes other modules and expresses local decisions about, e.g., which tools are differentially private.}\label{toplevelmodule}  
\end{figure*}
\end{document}